\documentclass[article, aps,prx,twocolumn,longbibliography,superscriptaddress]{revtex4-2}

\usepackage{graphicx}
\usepackage{dcolumn}
\usepackage{bm}
\usepackage{amsmath}
\usepackage{mathtools}
\usepackage{amssymb}
\usepackage{enumerate}
\usepackage{import}
\usepackage{xcolor}
\usepackage[normalem]{ulem}

\usepackage[colorlinks = true,
            linkcolor = blue,
            urlcolor  = blue,
            citecolor = blue,
            anchorcolor = blue]{hyperref}

\usepackage{orcidlink}

\newcommand{\la}{\langle}
\newcommand{\ra}{\rangle}
\newcommand{\ii}{\mathrm{i}}

\begin{document}

\title{Prolonging a discrete time crystal by quantum-classical feedback}

\author{Gonzalo Camacho\orcidlink{0000-0001-6900-8850}}%
\email{gonzalo.camacho@dlr.de}
 \affiliation{%
Institute for Software Technology, German Aerospace Center (DLR), Linder Höhe, 51147, Cologne, Germany
}
\author{Benedikt Fauseweh\orcidlink{0000-0002-4861-7101} } %
\email{benedikt.fauseweh@tu-dortmund.de}
 \affiliation{%
Institute for Software Technology, German Aerospace Center (DLR), Linder Höhe, 51147, Cologne, Germany
}
 \affiliation{
Department of Physics, TU Dortmund University, Otto-Hahn-Str. 4, 44227, Dortmund, Germany}

\date{\today}

\begin{abstract}
Non-equilibrium phases of quantum matter featuring time crystalline eigenstate order have been realized recently on noisy intermediate-scale quantum (NISQ) devices. While ideal quantum time crystals exhibit collective subharmonic oscillations and spatio-temporal long-range order persisting for infinite times, the decoherence time of current NISQ devices sets a natural limit to the survival of these phases, restricting their observation to a shallow quantum circuit. Here we propose a time-periodic scheme that leverages quantum-classical feedback protocols in subregions of the system to enhance a time crystal signal significantly exceeding the decoherence time of the device. As a case of study, we demonstrate the survival of the many-body localized discrete time crystal phase (MBL-DTC) in the one dimensional periodically kicked Ising model, accounting for decoherence of the system with an environment. Based on classical simulation of quantum circuit realizations we find that this approach is suitable for implementation on existing quantum hardware and presents a prospective path to simulate complex quantum many-body dynamics that transcend the low depth limit of current digital quantum computers.
\end{abstract}

\maketitle
\section{Introduction}

\begin{figure*}[!t]
\centering
\includegraphics[scale=0.42]{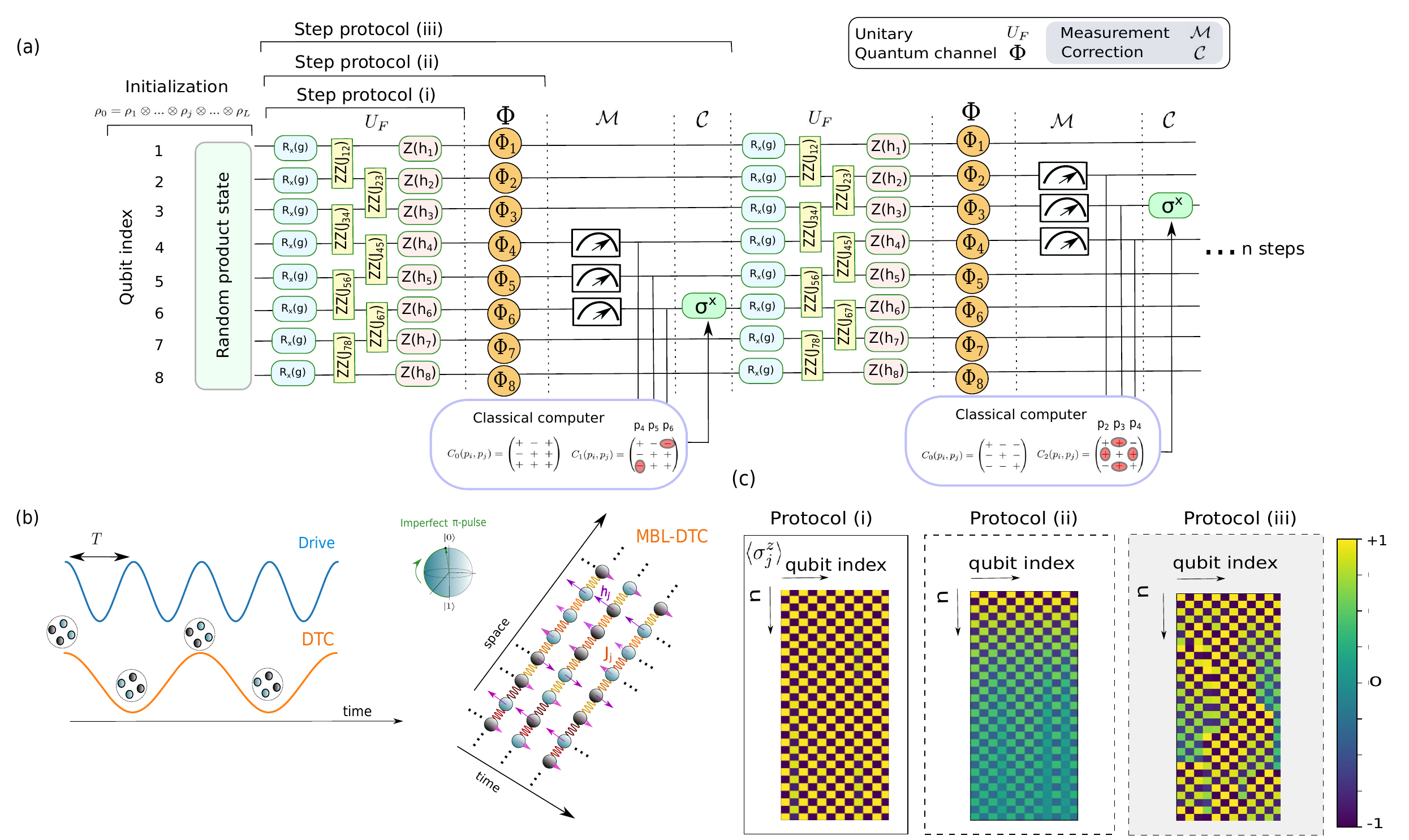}
\caption{Quantum-classical feedback scheme for the stabilization of an MBL-DTC on a quantum computer. (a) Quantum circuit realization for a chain of $L=8$ qubits. At initialization, an arbitrary product state with factorized density matrix $\rho_0$ is prepared in the computational basis. The evolution of the state is studied under three different protocols (i), (ii), (iii), with each protocol step containing different layers. The layers are applied in a periodic fashion during $n$ steps (circuit depth). We distinguish three different layers: a unitary layer $U_F$, a dissipative layer in the form of a quantum channel $\Phi$, and a non-unitary layer composed of a measurement $\mathcal{M}$ and a correction $\mathcal{C}$ operation, highlighted in grey background. For the measurement operation $\mathcal{M}$, a domain wall of $M$ qubits at an arbitrary position along the chain is measured in the $z$-basis; $M=3$ in the example. Measurement outcomes are recorded and processed classically to compute the correlation matrix between all measured qubits $C_n(i,j)$, which is compared with its value at initialization $C_0(i,j)$ for the same subset of qubits forming the domain wall. The qubit experiencing the most sign flips is identified and corrected. (b) A discrete time crystal (DTC) responds sub-harmonically in the presence of a external drive. The DTC can be regarded as an infinitely extended spatio-temporal lattice with period-doubled oscillations along the time dimension, with arbitrary couplings $J_j$ and local fields $h_j$ in space (different colors represent different coupling and field strengths). (c) The three different time evolution protocols: (i) A closed quantum system; (ii) An open quantum system; and (iii) An open quantum system subjected to a measurement and a correction operation $\mathcal{M}$ and $\mathcal{C}$, respectively. In the panels, we show the time evolution for the $z$-projection magnetization of the qubits $\langle \sigma_j^z\rangle$, for a single realization of the circuit with $L=12,M=3$ in (a) under protocols (i), (ii) and (iii).}
\label{fig:fig1}
\end{figure*}

\begin{figure*}[!t]
\centering
\includegraphics[scale=0.62]{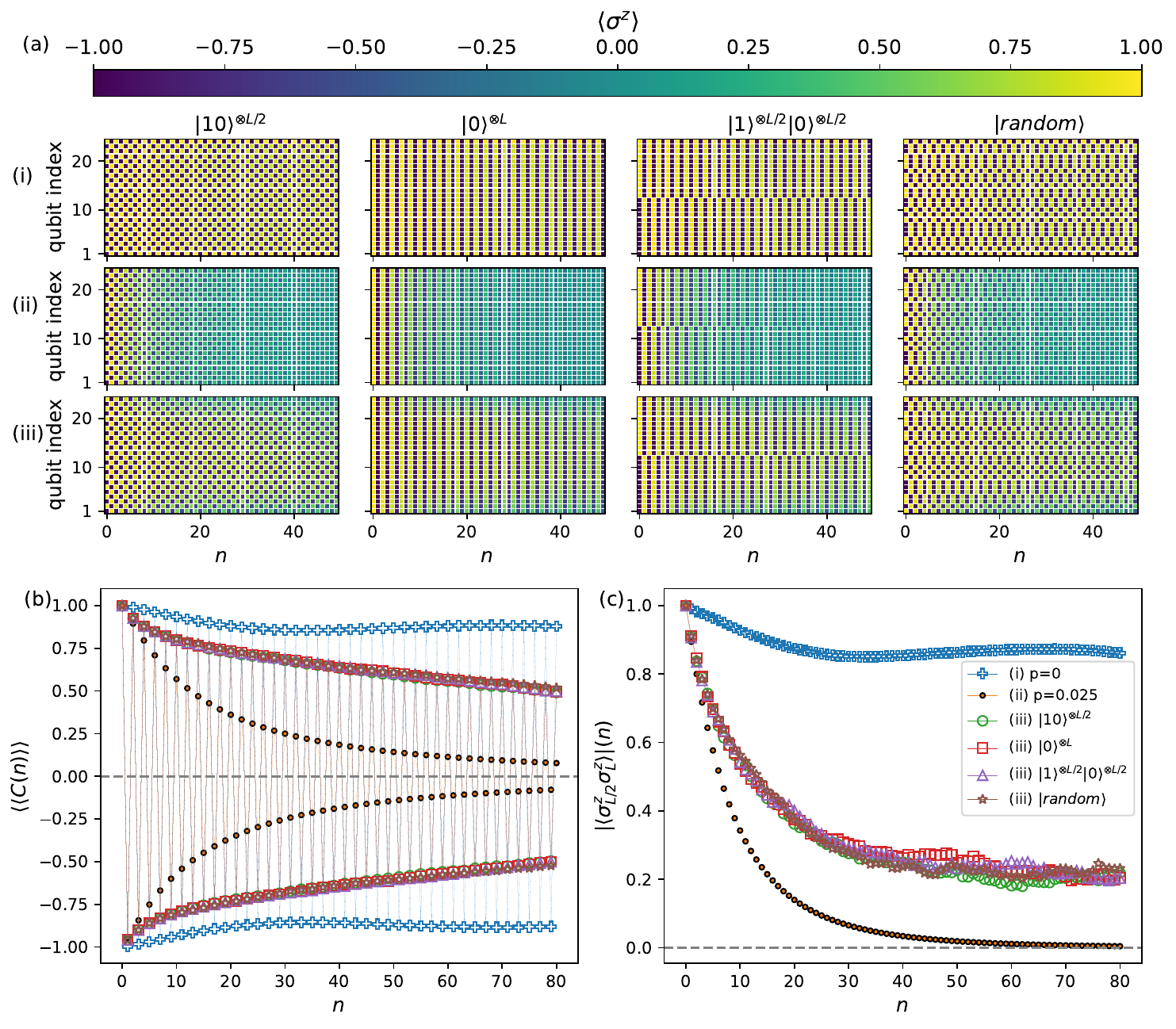}
\caption{Time evolution simulation for different initial product states. The system size is fixed to $L=24$ sites, $g=0.97$, $p=0.025$ (protocols (ii) and (iii)), employing a total of 80 steps on each protocol, and results are averaged over $N_{\text{dis}}=1280$ disorder realizations in all cases, with (b) and (c) only showing data for the initial state $|10\rangle^{\otimes L/2}$. The domain wall size in protocol (iii) is fixed to $M=6$ qubits, where at most one of them experiences a correction (see Fig.~\ref{fig:fig1} (a) and Appendix sections~\ref{subsubsec:measurement_prot} and~\ref{subsubsec:correction_protocol} for details). (a) The disorder averaged magnetization for the four different initial states (columns), for the three different time evolution protocols (rows), for different qubit positions and time steps $n$. (b) Disorder and spatially averaged autocorrelation. Legends correspond to those in (c). (c) The qubit-qubit correlations along the $z$-projection, between the qubit in the middle of the chain and that located at the right-most edge. }
\label{fig:fig2}
\end{figure*}

The exploration of non-equilibrium phases of matter has become a central focus in current many-body physics research, attracting significant attention and driving numerous research endeavors~\cite{Polkovnikov2011,Basov2017}. The fast development in the field has been motivated by the stunning degree of control achieved in experimental settings involving trapped ions~\cite{Leibfried2003,Blatt2012}, ultra-cold atoms~\cite{Bloch2012,Gross2017}, light-induced superconductors~\cite{Fausti2011} and ultrafast topology switching \cite{Sie2019}, to name a few. More recently, the growing interest on making use of available noisy-intermediate scale quantum (NISQ) devices~\cite{Preskill2018} has opened the path for intensive research in the field of digital quantum simulation applied to complex non-equilibrium phenomena.

Understanding the capabilities of current digital quantum simulators for realizing non-equilibrium phases of matter is a problem on demand~\cite{Smith2019,Fauseweh2021,Zhang2022,Kamakari2022,Fisher2023,Fauseweh2024}. It is widely accepted that NISQ devices suffer from a major limitation to perform quantum computations due to their intrinsic decoherence times. In digital quantum simulation, where the study of many-body quantum dynamics is based on generic Trotterization schemes, this translates into a pronounced loss of fidelity as the number of gates composing a quantum circuit is increased~\cite{Lanyon2011,Salathe2015}. These findings naturally lead to questions about the potential of NISQ machines beyond the constraints of low circuit depth, particularly in terms of simulating quantum many-body systems and detecting emergent phases of matter that are out of equilibrium.

Recent experiments performed on quantum computers suggest that error mitigation techniques can successfully push the applicability of current NISQ devices to perform quantum simulations~\cite{Kim2023}, while the long-term goal for achieving full quantum error correction~\cite{Shor1995} is still under investigation. In parallel, research has been conducted in the field of monitored quantum dynamics~\cite{Fisher2023} to understand how the application of local measurement operations affects the evolution of a quantum system, particularly concerning its entanglement dynamics~\cite{Vasseur2019,Skinner2019,Gullans2020,Bao2020,Ippoliti2021prx,Nahum2021,Jian2021,Ippoliti2021prl,Lavasani2021,Weinstein2022,Sierant2023,Nandkishore2023,Morral2024}. A less explored approach in this domain is to employ the outcomes obtained from measurements to construct feedback protocols that steer the evolution in a desired direction. Such quantum-classical feedback schemes are beginning to be considered in superconducting qubit platforms as a way to improve current qubit fidelities~\cite{Vepsaelaeinen2022}. Along these lines, we pose the following question: can we use in-circuit quantum-classical feedback protocols to simulate quantum many-body dynamics beyond the intrinsic decoherence time of the device?

To address this question, we look at an example of a genuine non-equilibrium phase of matter implementable on a quantum device~\cite{PRXQuantumSondhi2021}; the quantum time-crystals~\cite{Wilczek2012,Sacha2017,khemani2019brief,zaletel2023}. These systems have been observed recently in experiments to feature spontaneous symmetry breaking of time translation either in a continuous~\cite{Kongkhambut2022,Liu2023} or in a discrete fashion~\cite{Zhang2017,Choi2017,Krzysztof2018,DTCexperiment1,DTCexperiment2,Bernien2017,Ho2017,Pal2018,Smits2018,OSullivan2020,Kessler2020a,Kessler2021,Kleiner2021,Taheri2022,Autti2022,Taheri2022a,Bluvstein2021,Hannaford2022}; the discrete case corresponding to the so called Discrete Time Crystals (DTC)~\cite{Else2016,Khemani2016,Moessner2017,Yao2017,Hess2017,Else2020}. DTC manifest emergent spatio-temporal long-range order, and an overall sub-harmonic response with infinitely long-lived oscillations~\cite{khemani2019brief,zaletel2023,Else2016,Khemani2016,Moessner2017,Yao2017,Hess2017,Else2020,Iemini2018,Huang2018,Mizuta2018,Pizzi2019,Pizzi2021,Pizzi2021a,Maskara2021,Collura2022,Alaeian2022,McGinley2022,Huang2022,Sacha2015,Nurwantoro2019,Kuros2020,Yu2019,Fan2020,Li2020,Wang2021,Gong2018,Krishna2023,Liu2023prl} in the presence of an external drive. In disordered and interacting systems, the combination of many-body localization (MBL)~\cite{Gornyi2005,Basko2006,Nandkishore2015,Abanin2019} and DTC has resulted in the emergence of a genuine non-equilibrium phase of matter in the framework of eigenstate order~\cite{Huse2013,Abanin2015,Keyserlingk2016,Moessner2017}; the resulting phase has been termed a many-body localized discrete time crystal (MBL-DTC)~\cite{Khemani2016,Else2016,PRXQuantumSondhi2021}. The existence of the MBL-DTC has been supported by experiment~\cite{Randall2021} and its recent realization on NISQ hardware~\cite{Mi2022,Frey2022}. However, due to decoherence effects experienced by the qubits forming a quantum circuit, the MBL-DTC has been observed only for a few number of cycles on NISQ computers. It is indeed expected that when the system is coupled to a Markovian environment, the overall subharmonic response characterizing the DTC phase gets eventually lost. Understanding the general conditions under which DTC phases might survive in open quantum systems is still a subject of intensive research~\cite{Lazarides2017,Buca2019,Gambetta2019prl,RieraCampeny2020,Sarkar2022,Chinzei2022,Sankar2023}.

In this work, we propose a stabilization protocol directly implementable in quantum circuits, that combines measurements on the system with a classical error correction code. The protocol is based on the periodic application of projective measurements in randomly located domain walls across subregions of the system, with the in-circuit classical computation of correlations taking place after a measurement event. More explicitly, we employ a correction scheme that identifies single qubit flips based on the construction of the correlation matrix of measurement outcomes and the comparison with its value at initialization. The outcome obtained from the classical processing is then employed to determine the target qubit undergoing a correction operation (see Fig.~\ref{fig:fig1} (a) and Appendix section~\ref{subsubsec:correction_protocol} for details on the specific correction protocol employed).

We focus on the particular case of the MBL-DTC phase recently observed in Ref.~\cite{Mi2022}, and whose model is schematically represented in Fig.~\ref{fig:fig1} (b). We test and verify our scheme employing classical simulation of quantum circuits realizations, comparing the unitary protocol (i), the open quantum system (ii) and the stabilization protocol in the open system (iii), all of which are represented in Fig.~\ref{fig:fig1} (c).

\section{Model and protocol}\label{sec:Model}

As a typical example for an MBL-DTC we consider the one dimensional periodically kicked Ising model. The model consists of a one dimensional chain of $L$ qubits with open boundary conditions and nearest-neighbour interactions, subjected to an external periodic drive of period $T$. The Floquet unitary operator has the form~\cite{Khemani2016,PRXQuantumSondhi2021,Mi2022,Frey2022}:
\begin{eqnarray}\label{eq:floquet_op}
U_F=e^{-\ii\frac{T}{4}\sum_{j=1}^{L-1} J_j\sigma_j^z\sigma_{j+1}^{z}}e^{-\ii\frac{T}{2}\sum_{j=1}^{L}h_j\sigma^{z}_j}e^{-\ii \frac{\pi g}{2}T\sum_{j=1}^L\sigma^{x}_{j}},\nonumber\\
\end{eqnarray}
where $\sigma_j^{x},\sigma_j^{z}$ are the $x,z$ Pauli matrices at site $j$ of the chain, respectively. The parameter $T$ represents the Floquet period of the external drive, which we set to $T=1$. The parameter $g$ is the pulse parameter, with $g=1$ representing a perfect $\pi$-pulse on the Bloch sphere. The coupling parameters $J_j$ and the magnetic fields $h_j$ are randomly sampled from uniform distributions~\cite{Mi2022} with $J_j\in\left[-1.5\pi,-0.5\pi\right]$ and $h_j\in[-\pi,\pi]$. With this choice of the parameters, the model in Eq.~\eqref{eq:floquet_op} is exactly equivalent to the one in Ref.~\cite{Mi2022}, and has been implemented recently on NISQ hardware~\cite{Frey2022,Mi2022}. The model has three different phases (paramagnetic, thermal, MBL-DTC) depending on the values of $g$~\cite{PRXQuantumSondhi2021}; here we focus on the MBL-DTC phase of the model, which has been reported to exist for values of $g\gtrapprox 0.82$~\cite{PRXQuantumSondhi2021,Mi2022}.

Protocol (i) corresponds to unitary evolution of the state in a closed quantum system. For protocols (ii) and (iii), we assume that the system is open and coupled to a dissipative Markovian environment modelled as a quantum channel $\Phi$ depending on a single noise parameter $p$ (see Appendix section~\ref{subsec:Noise_models} for details on the noise model employed). The coherent errors that might take place on a real setup can generally be mapped to incoherent error models through the Kraus formalism, in the spirit of Ref.~\cite{Wallman2016}. We thus assume here an incoherent noise model, noting that the application of more sophisticated and realistic noise models are possible by employing such approaches. All time evolution protocols are carried out employing tensor networks and by representing the state of the system as a Matrix Product Density Operator (MPDO), see Appendix sections~\ref{subsec:mpdo},~\ref{subsec:numconv} and~\ref{subsec:edbench} for details. Due to the existence of an infinite number of local conservation laws~\cite{Nandkishore2015,Serbyn2013b}, closed systems featuring MBL effects generically obey an unbounded logarithmic growth of the entanglement entropy~\cite{Bardason2012,Serbyn2013} in the thermodynamic limit. For a one-dimensional system, such area-law behavior of entanglement translates into an efficient representation of these systems employing Matrix Product State (MPS) methods~\cite{Friesdorf2015}. 

\section{Results}\label{sec:Results}

\subsection{Stabilization for different initial states}
\label{subsec:init_state_indepe}
Characterizing a true MBL-DTC phase requires independent results regardless of the choice of any initial product state~\cite{khemani2019brief,PRXQuantumSondhi2021}. Here we study the time evolution of the model in Eq.~\eqref{eq:floquet_op} under protocols (i), (ii) and (iii), for a set of different initially prepared product states: the N{é}el state $|10\rangle^{\otimes L/2}$, the zeros state $|0\rangle^{\otimes L}$, the wall state $|1\ra^{\otimes L/2}|0\ra^{\otimes L/2}$, and a random bit-string $|\text{random}\rangle$.

\begin{figure}[!t]
\includegraphics[scale=0.7]{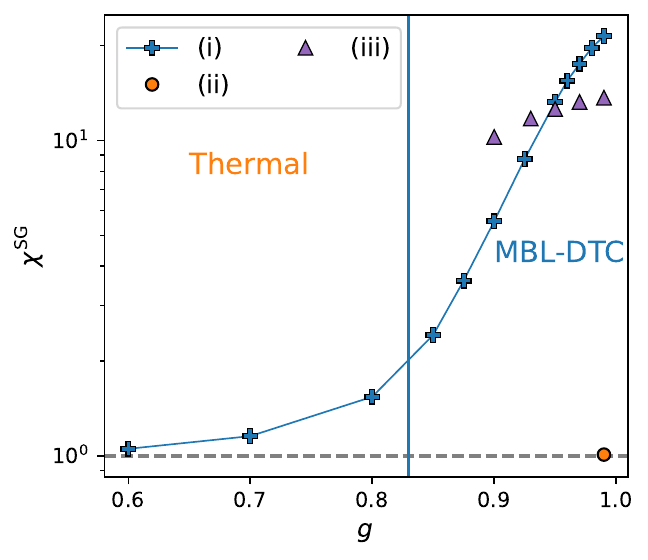}
\caption{Glassy spatial order enhancement. The spin glass order parameter $\chi^{\text{SG}}$ as a function of the pulse parameter $g$, obtained by averaging over random initial product states and disorder realizations for $L=24$ qubits, and $p=0.025$ for protocols (ii) and (iii). We generate a total of $N_{\text{dis}}=1280$ disorder realizations for all protocols. Following Ref.~\cite{Mi2022}, we employ a total of 60 time steps. $\chi^{\text{SG}}$ is obtained by averaging the values between steps 48 and 60. The horizontal dashed line corresponds to $\chi^{\text{SG}}=1$. The straight line joining data points in protocol (i) has been included as a guide to the eye.}
\label{fig:fig3}
\end{figure}

Our first observation is that results are independent of the initial preparation of the state. In Fig.~\ref{fig:fig2} (a), we have represented the disorder averaged magnetization for the four product states considered. The application of protocol (iii) leads to an overall restoring of the MBL-DTC phase in all cases, which is otherwise rapidly lost under protocol (ii) due to repeated action of the quantum channel~\cite{PRXQuantumSondhi2021} at each step $n$. This is our main result: The correction scheme enhances the subharmonic response beyond the intrinsic decoherence time of the device. We note that with the proposed correction scheme in protocol (iii), qubits located at the edges of any finite chain have a lower probability of correction than qubits deep in the bulk, an effect that is appreciated in Fig.~\ref{fig:fig2} (a). This effect is due to the open boundary conditions employed in the correction protocol (iii), for which an edge qubit has a probability $1/(L-M)$ of being part of a domain wall. On the other hand, bulk qubits have a probability of $M/(L-M)$ of belonging to a randomly located domain wall, and thus of being corrected. This effect is expected to be absent for bigger system sizes, i.e. when addressing the thermodynamic limit.

In order to quantify the effect of the correction scheme  onto the subharmonic response, we show in Fig.~\ref{fig:fig2} (b) the spatially and disorder averaged autocorrelation $\langle\langle C(n)\rangle\rangle$ (see Appendix section~\ref{subsec:observables}). Employing protocol (iii)  greatly enhances the sub-harmonic response of the system when compared to the pure dissipative case of protocol (ii). The qubit-qubit correlations between the mid-chain qubit and the right-most qubit of the chain have been represented in Fig.~\ref{fig:fig2} (c). The results for protocol (iii)  show a saturation to a finite value for increasing number of steps, a behavior in accordance with the unitary protocol (i), independent of the initial state.

To further investigate the robustness of the results with respect to different initial (product states) preparations, we proceed to calculate the Edwards-Anderson spin-glass order parameter $\chi^{\text{SG}}$. This extensive quantity identifies the amount of spatial spin-glass order developing in the system. As the initial condition, we prepare a set of random product states in the computational basis, keeping the same values for $L, p$ and $M$ in all cases.

The spin glass order parameter is represented in Fig.~\ref{fig:fig3}. In accordance with previous results~\cite{Mi2022}, protocol (i) shows a clear transition from the thermal to the MBL-DTC phase of model Eq.~\eqref{eq:floquet_op}, when averaged over different initial (short-ranged) states and different disorder realizations. Under protocol (ii), glassy spatial order is absent for the considered noise model due to decoherence, showing a value $\chi^{\text{SG}}\sim 1$. The application of protocol (iii) shows a restoring of the $\chi^{\text{SG}}$ order parameter to values indicating a high degree of spatial correlation between qubits, i.e. the appearance of glassy spatial patterns as in the unitary case of protocol (i). Tuning the value of $g$ towards the thermal phase becomes computationally expensive due to the growth of entanglement; thus, we limit ourselves to results with $g\geq 0.9$.

\begin{figure*}[!t]
\includegraphics[scale=0.39]{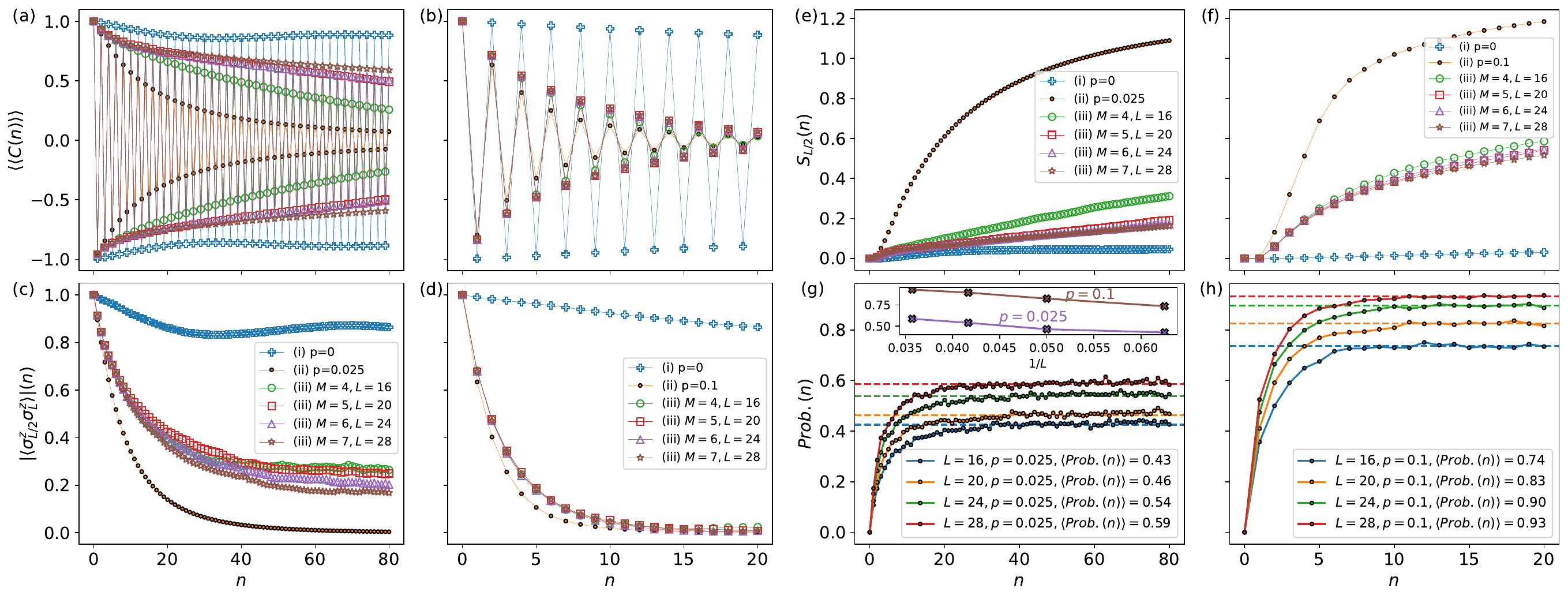}
\caption{System size scaling and noise dependence. Comparison of the different time evolution protocols starting from the initial state $|10\rangle^{\otimes L/2}$, for $g=0.97$ and $p=0.025,0.1$, and different system sizes $L$. All results were obtained averaging over $N_{\text{dis}}=5120$ disorder realizations. Straight lines between data points in all figures are included as a guide to the eye. Panels (a),(b): The spatially and disordered averaged autocorrelation at stroboscopic times. Panels (c),(d): The qubit-qubit correlations between the mid-chain qubit and the qubit located at the right-most edge of the chain. Panels (e),(f): The mid-chain operator entanglement entropy (see Appendix section~\ref{subsec:observables}). Under protocol (iii), the measurement protocol operations lead to an overall slow growth of the operator entanglement entropy.  Panels (g), (h): Probability of performing a single qubit correction under protocol (iii), for different system sizes and noise levels. The inset in (g) shows the values marked by the dashed lines as a function of $1/L$, for the two noise levels considered. }
\label{fig:fig4}
\end{figure*}

\subsection{Scaling and Relevance of the correction scheme}
A key question to address concerns the scaling of protocol (iii) with increasing system size $L$ and noise level $p$, as well as the verification of the relevant role played by the correction scheme in the observed behavior. Hence, we prepare the initial state to $|10\rangle^{\otimes L/2}$, studying variations of the $L$ and $p$ parameters, as well as a different number of qubits undergoing a correction operation.

In Figs.~\ref{fig:fig4} (a),(b) ((c),(d)), we represent the autocorrelation (the qubit-qubit correlation) for two different noise levels. With increasing system size, we observe that for moderate noise levels $p=0.025$, employing the correction protocol leads to robust sub-harmonic response and stabilization of correlations compared to protocol (ii). The enhancement of correlations in the $p=0.025$ case is a consequence of the specific design of the correction scheme employed in protocol (iii), which relies on the classical processing of the correlation matrix between qubits forming the domain wall. As the correlation matrix increases its size, the identification of a defective qubit based on the total number of sign flips becomes more accurate (see Appendix section~\ref{subsubsec:correction_protocol} for details on the correction protocol). The finite value of correlations between qubits indicates a restoring of spatial order. For strong noise $p=0.1$, we observe that protocol (iii) is incapable of overcoming noise, and the system evolves very close to the dissipative case in protocol (ii).

\begin{figure*}[!t]
\includegraphics[scale=0.70]{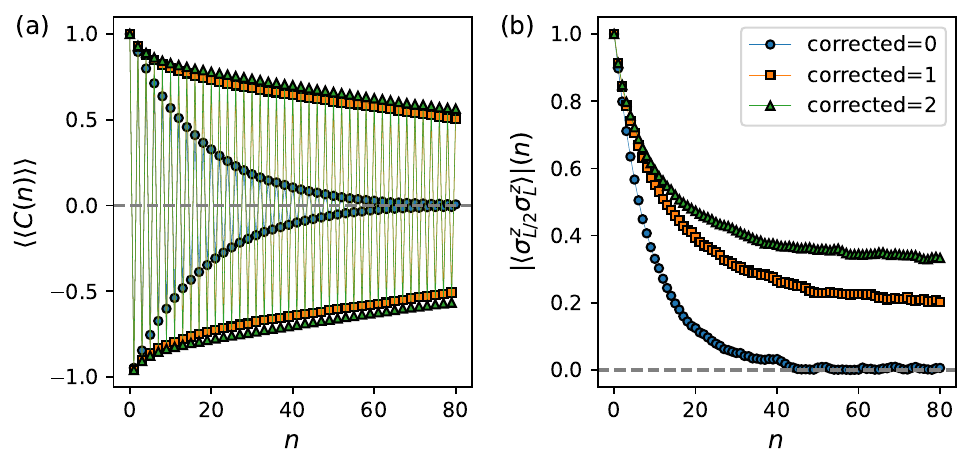}
\caption{Verification of enhancement due to correction scheme: autocorrelation (a) and qubit-qubit correlations (b) for $L=24$, $g=0.97$, $p=0.025$, $M=6$ and $N_{\text{dis}}=5120$ samples, for protocol (iii) when a different number of qubits are corrected under the operation $\mathcal{C}$. The absence of the correction protocol $\mathcal{C}$ (i.e. the case with zero corrected qubits) does not lead to an enhancement of sub-harmonic response, with both spatial and temporal correlations showing a rapid decay.   }
\label{fig:fig5}
\end{figure*}

In Figs.~\ref{fig:fig4} (e),(f), we observe a pronounced growth of the mid-chain operator entanglement entropy (see Appendix section~\ref{subsec:observables} for the definition of this quantity) for the two noise levels of $p$ under protocol (ii). The application of protocol (iii) leads to an overall slow growth in the dynamics of entanglement, with the case $p=0.025$ showing closer behavior to the unitary case. We note that the employed correction scheme does not lead to a complete saturation of entanglement in any case. In Figs.~\ref{fig:fig4} (g),(h) we include statistics on the single-qubit probability of correction under protocol (iii). Increasing the value of the noise level leads to a higher probability of correction, with all curves saturating to finite values in the long-time limit. The relation between saturation values for the correction probability and system size for both noise levels is included in the inset in Fig.~\ref{fig:fig4} (g). For moderate values of $p$, the probability eventually stabilizes to a finite value dependent on the system size $L$, with bigger $L$ showing higher correction probabilities, as expected due to the increased possibility of detecting a bit-flip event after measuring bigger domain walls.

The results indicate that the correction scheme in protocol (iii) is able to correct to a certain degree erroneous outcomes of the qubits experiencing a measurement operation $\mathcal{M}$. In the context of monitored quantum dynamics in random circuits with repeated local measurements, a field where quantum-classical feedback remains uncharted ~\cite{Fisher2023}, we now seek to address whether the measurement operations in $\mathcal{M}$  alone can restore the  subharmonic response without the necessity of a correction scheme. To further verify that the restoring of the DTC behavior takes place due to the applied correction scheme, we have represented in Fig.~\ref{fig:fig5} the autocorrelation and the qubit-qubit correlations for protocol (iii) employing a different number of corrected qubits in $\mathcal{C}$. We observe that the absence of a correction protocol (zero corrected qubits) is not enough to restore the sub-harmonic response of the system, i.e. local measurement operations $\mathcal{M}$ alone do not on average preserve the MBL-DTC phase. Moreover, we observe that the repeated application of measurements alone lead to a slightly faster decay than in the pure dissipative case of protocol (ii), compare Fig.~\ref{fig:fig4} (a). In contrast, the composed operation $\mathcal{C}\circ \mathcal{M}$ with at least one corrected qubit in $\mathcal{C}$ is able to partially restore the subharmonic response of the system. We conclude that the correction scheme employed here outperforms a pure Zeno effect in the presence of dissipation.

\subsection{Long time limit}
Here we address the question of the survival of the MBL-DTC under protocol (iii) in the large circuit depth limit, under the tuning of several parameters, namely the system size, the noise level and the pulse parameter.

In Figs.~\ref{fig:fig6} (a),(b) we have represented the absolute value for both the autocorrelation and the qubit-qubit correlation function for different system sizes and domain wall sizes. We observe that under protocol (iii), there is a pronounced enhancement of the subharmonic response of the system, with a much slower decay in $\langle\langle C(n)\rangle\rangle$ with increasing value of the domain wall $M$, confirming that the proposed correction protocol is able to accurately identify qubit flips that destroy the subharmonic pattern following a local measurement operation for a noise level $p=0.01$.   

In Figs.~\ref{fig:fig6} (c),(d) we report results for the long-time behaviour under variations of the parameters $p$ and $g$, fixing the system and domain wall sizes. For a noise level of $p=0.01$, protocol (iii) is able to restore spatio-temporal order for times way beyond the decoherence scale in the model, with increasing values of $p$ posing a strong limitation in the application of the correction protocol. The tuning of $g$ towards values approaching the thermal region plays a minor role compared to the effect of noise. 

The results in Figs.~\ref{fig:fig6} indicate that even in the presence of small noise, the protocol approaches the classical limit at the point $g=1$, where perfect subharmonic response is expected, but no quantum entanglement develops in the system. This occurs as a consequence of the application of local projective measurements in protocol (iii), leading to an overall decrease of entanglement of the state evolution, while at the same time, increasing the purity.

\section{Discussion}
\label{sec:summary}

\begin{figure*}[!t]
\includegraphics[scale=0.7]{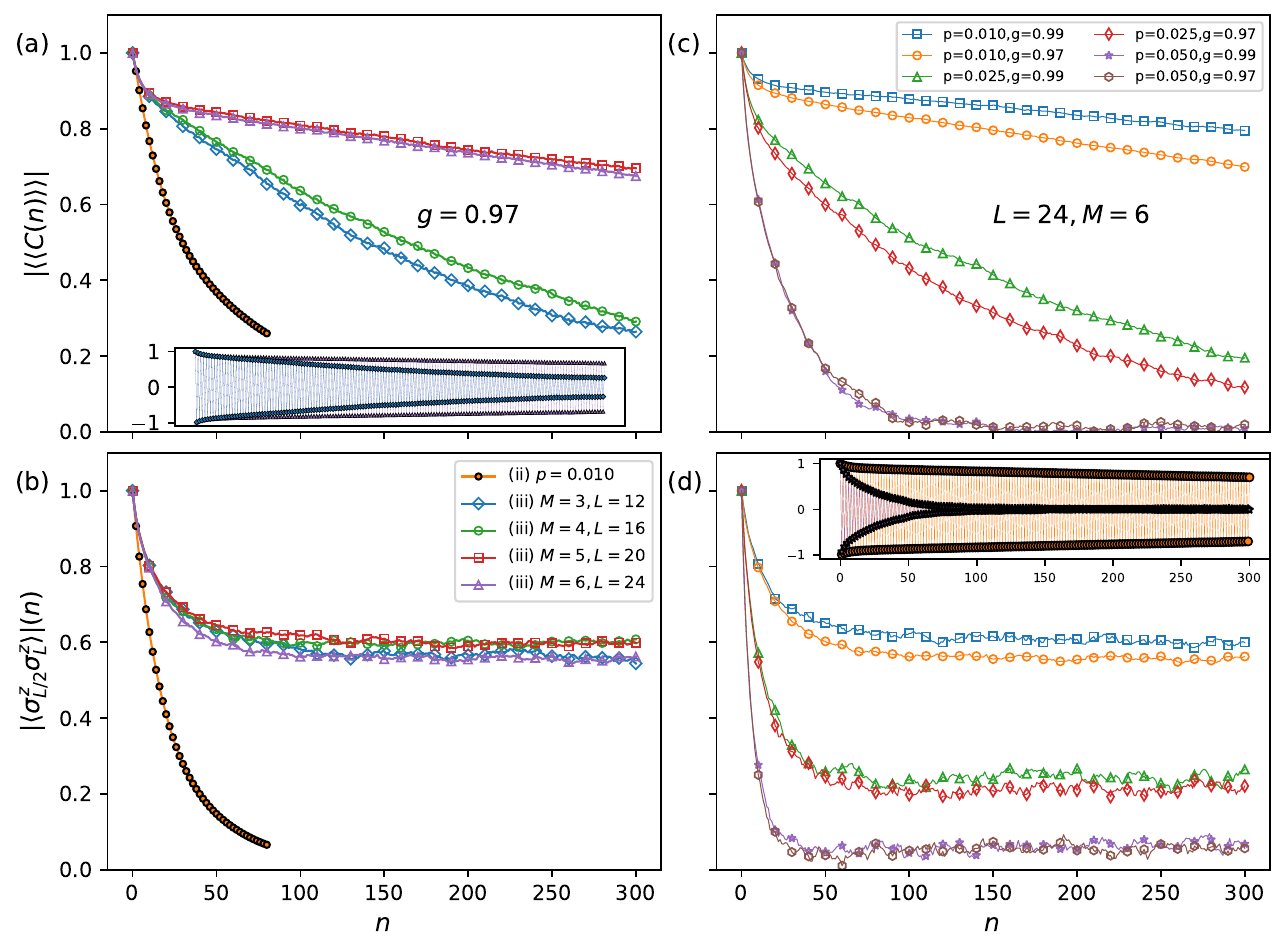}
\caption{Long time limit under different tuning of parameters. Panels (a),(b): Absolute value of (a) the autocorrelation and (b) the qubit-qubit correlation function, for different $L,M$, $g=0.97$, $p=0.01$ and a total of $300$ time steps. Data corresponding to protocol (iii) was averaged over $N_{\text{dis}}=1280$ disorder realizations; for protocol (ii), results are averaged over $N_{\text{dis}}=256$ for 80 time steps and $L=24$. For protocol (iii), data markers have been included every ten time steps. The inset in (a) shows the subharmonic oscillations obtained for $\langle\langle C(n)\rangle\rangle$ in the cases $M=3$ (orange) and $M=6$ (purple) under protocol (iii). Legend labels in (b) apply to (a). Panels (c),(d): Absolute value of (c) the autocorrelation and (d) the qubit correlation function under protocol (iii), for $L=24,M=6$, averaging over $N_{\text{dis}}=1280$ disorder realizations, for different values of $g$ and $p$. Data markers have been included every ten steps. The inset in (d) shows the subharmonic response obtained in the autocorrelation in (c), for the cases $p=0.05,g=0.99$ (purple) and $p=0.01,g=0.97$ (orange). Legend labels in (c) apply to (d). }
\label{fig:fig6}
\end{figure*}

In this work, we have proposed a scheme to overcome dissipation effects in the MBL-DTC phase recently observed in NISQ devices~\cite{Frey2022,Mi2022}. Our approach is based on the application of a periodic quantum-classical feedback protocol, where single qubit correction based on classical processing of measurement outcomes is able to enhance spatio-temporal order, eventually restoring the discrete time crystal signal well beyond the decoherence time of the device. We demonstrate the capabilities of the protocol by systematically investigating the dependence on initial state, system size and noise level, while addressing the question of how do feedback schemes based on local projective measurements affect the monitored dynamics of open quantum systems~\cite{Fisher2023}. 

The increase in the subharmonic response with increasing system size, Fig.~\ref{fig:fig6}, suggests a complete stabilization of the MBL-DTC in the thermodynamic limit. This can not be proven rigorously with the system sizes we can simulate classically. Demonstrating the stabilization effect on a real NISQ device therefore presents a rare opportunity to obtain a deeper understanding about quantum many-body dynamics from a digital quantum computer.

There are several directions and extensions that can be explored using similar stabilizing schemes. From the theory perspective of DTC, we motivate further studies that employ more sophisticated noise models, different measurement schemes and alternative classical (or quantum) error correction protocols, all in the context of how spatio-temporal order gets affected by such local controlled operations. We believe that a particularly interesting question to address concerns the extension of the method to multi-qubit correction schemes employing more than a single domain wall in the measurement process. Providing further insight from a fundamental point of view in order to better understand the observed collective behaviour is also highly desirable, particularly regarding the general conditions necessary for the observation of discrete time crystals in more general open quantum systems~\cite{Lazarides2017,zaletel2023}. In this case achieving full control over the phase is essential for developing efficient and long-lived quantum memory devices. 

Due to its conceptual design the approach is suitable for immediate implementation on current quantum hardware. The correction protocol discussed in this work needs sufficient idle times in the qubits in order to perform the readout and classical processing following a projective measurement. We note that such functionality is at the heart of quantum error correction schemes, constituting a very active area of research regarding improved coherence qubit times~\cite{Vepsaelaeinen2022}. In  superconducting qubit platforms \cite{baumer2023efficient} as well as trapped ion platforms \cite{PhysRevX.13.041052}, employing feedback schemes is now possible, making our protocol feasible in state-of-the-art devices.

A potential application concerns the stabilization of phases of quantum matter in digital quantum devices, where the use of current NISQ devices finds strong limitations due to decoherence processes. An interesting direction concerns the investigation of related in-circuit feedback protocols in the simulation of diverse many-body quantum systems, in particular regarding  the study of recently realized measurement-induced quantum phases in trapped ion quantum computers~\cite{Noel2022}. We highlight the potential application of these methods in simulating quantum systems defined in higher dimensional spaces, with qudit-based processors~\cite{Ringbauer2022} standing out as candidates where such stabilization protocols could lead to the observation of novel out of equilibrium phases of quantum matter.   

\section*{Data availability}

The data supporting the findings of this study are available from Zenodo at \url{https://doi.org/10.5281/zenodo.8318707} and also upon request from the corresponding author. 

\section*{Acknowledgments}

This project was made possible by the DLR Quantum Computing Initiative and the Federal Ministry for Economic Affairs and Climate Action of Germany; \url{qci.dlr.de/projects/ALQU}.

\appendix

\section{}

\subsection{Matrix Product Density Operators (MPDO)}
\label{subsec:mpdo}
We represent the state of the system $\rho$ in terms of a finite number of variational parameters, in the form of a Matrix Product Density Operator~\cite{Verstraete2004,Schollwoeck2011} (MPDO):
\begin{eqnarray}\label{eq:mpdo}
\rho &=& \sum_{\sigma_1,\sigma_1',...,\sigma_N,\sigma_N'} M^{\sigma_1\sigma_1'}...M^{\sigma_N\sigma_N'},\nonumber\\
M^{\sigma_j\sigma_j'}_{D_jD_{j+1}}&=&\sum_{r=1}^{K} A^{\sigma_j,r}_{\chi_j\chi_{j+1}}\otimes \bar{A}^{\sigma_j',r}_{\chi'_j\chi'_{j+1}},
\end{eqnarray}
where the matrices $M^{\sigma_{1}\sigma_{1'}}_{D_jD_{j+1}}$ have dimensions $D_j=\chi_j\chi'_j$ and $D_{j+1}=\chi_{j+1}\chi_{j+1}'$, with $r$ representing an auxiliary space index. To carry out the time evolution of the state, we employ the Time Evolving Block Decimation (TEBD) method~\cite{Vidal2003,ZwolakVidal2004,Schollwoeck2011,Paeckel2019,Orus2021} extended to open quantum systems, in the spirit of Ref.~\cite{Eisert2016}, where local updates of the tensor $\rho$ are performed employing a series of Singular Value Decompositions (SVD). The dimension $\chi$ plays the role of the bond dimension employed in the unitary evolution of Matrix Product States (MPS)~\cite{Schollwoeck2011}. The error upon SVD decompositions after application of unitary entangling gates is controlled by a TEBD tolerance parameter $\varepsilon$. We increase dynamically the value of the bond dimension $\chi$ whenever the tolerance $\varepsilon$ is exceeded after every SVD decomposition. 

The dimension $K$, which is identified with the Kraus dimension of the channel, generally grows as a result of subsequent SVD decompositions~\cite{Eisert2016,Cheng2021} performed in the auxiliary indices. Upon application of a non-unitary gate, only the lowest $K$ singular values are retained. Thus, there are two main sources of error, one associated with the growth of entanglement in the system, and one associated with the statistical noise error due to the channel employed. For the data displayed in the main text, we fix $\varepsilon=10^{-4},K=10$. We have checked that choosing smaller (larger) values of $\varepsilon$ ($K$) do not affect the results in a significant way (see Appendix section~\ref{subsec:numconv}). As a technical detail, projective measurement operations are realized as non-unitary local quantum gates, which are known to destroy the Schmidt decomposition in the standard TEBD method~\cite{Vidal2003}. This can be overcome by the application of identity operations sweeping across the chain, in analogy with the approach followed in the imaginary time evolution TEBD.

\subsection{State initialization and time evolution protocols}
\label{subsec:State_init}
Any initially prepared state is represented by a product state with factorized density matrix $\rho_0$ in the computational basis:
\begin{eqnarray}\label{eq:rho0}
\rho_0 &=& \rho_1\otimes...\otimes\rho_j\otimes...\otimes\rho_L,\nonumber\\
\rho_j &=& |0\ra\la 0| \lor |1\ra\la 1|.
\end{eqnarray}
To study disorder ensembles with different couplings in Eq.~\eqref{eq:floquet_op}, we generate a total of $N_{\text{dis}}$ disorder realizations, each realization corresponding to a different circuit of the form represented in Fig.~\ref{fig:fig1} (a) and having its own initial state $\rho_0$. We study each disorder realization independently, with results being averaged over all disordered samples in the end. Note that for a single realization of the coupling parameters, any measurements carried out in real quantum hardware would need to be repeated several times for those same values of the couplings, in order to reduce shot noise. 

We focus on the time evolution of the state $\rho$ at stroboscopic times $n$ under three different time evolution protocols schematically represented in Fig.(1) (c) in the main text.

In protocol (i), the system is closed, and the unitary evolution is carried out by direct action of the Floquet unitary $U_F$. In protocol (ii) the system is open, and we model decoherence effects in the system by applying a dissipation layer $\Phi$ immediately after the action of the unitary layer $U_F$, with each qubit experiencing a quantum channel. The full quantum channel acts over the state as a mapping $\Phi(\rho)$ chosen to be a completely positive and trace-preserving map (CPTP). Under protocol (iii), the dissipative layer is followed by the application of a non-unitary layer formed by a measurement operation $\mathcal{M}$ on a subset of the qubits, which is chosen to be a domain wall of size $M$ that is randomly located at any time step $n$. The second operation $\mathcal{C}$ consists of a (classical) correction protocol where the outcome of the measured qubits is processed in a classical way (see the correction operation section below). 

The three different time evolution protocols and their corresponding operations over the state of the system at a given time step $n$ are:
\begin{enumerate}[\emph{Protocol} (i)]
    \item Unitary evolution of the state:
    \begin{eqnarray}
        \rho_n=U_F^n \rho_0 (U_F^\dagger)^n.\nonumber
    \end{eqnarray}
    \item Unitary step followed by quantum channel $\Phi$:
    \begin{eqnarray}
        \rho_n = \Phi\left(U_F\rho_{n-1}U_F^\dagger\right).\nonumber 
    \end{eqnarray}
    \item Unitary step, followed by quantum channel $\Phi$, measurement $\mathcal{M}$ and a correction operation $\mathcal{C}$:
    \begin{eqnarray}
        \rho_n = \mathcal{C}\circ\mathcal{M}\circ\Phi\left(U_F\rho_{n-1}U_F^\dagger\right).\nonumber
    \end{eqnarray}    
\end{enumerate}
Note that the order of operations is taken from right to left, with each composed operation taking place at any step $n$.

\subsection{Dissipative layer}
\label{subsec:Noise_models}

We focus on arguably one of the simplest noise channel to represent the dissipative layer $\Phi$, namely the bit-flip channel. We consider that each of the qubits experiences a bit-flip channel immediately after application of the Floquet unitary (see Fig.(1) (a) in the main text). This noise model assumes only local decoherence effects in the qubits. For the DTC considered in this work, the assumption of local bit-flip effects is a natural choice; however, extensions of the method to more realistic noise models is straightforward. 

The quantum channel is described by a Completely Positive Trace Preserving (CPTP) map $\Phi$, whose action over the state $\rho$ is represented by a set of $Q$ Kraus operators $K_i$:
\begin{eqnarray}\label{eq:kraus_map}
\Phi(\rho)= \sum_{i=0}^{Q-1} K_i \rho K_i^\dagger, \hspace{10pt}\sum_i K_i^\dagger K_i =I,
\end{eqnarray}
where $I$ is the identity matrix. The single-qubit bit-flip channel depends on a single noise parameter $p$, and it is defined by two Kraus operators:
\begin{eqnarray}
K_0 = \sqrt{1-p}I,\hspace{5pt} K_1=\sqrt{p}\sigma^x.
\end{eqnarray}
This quantum channel corresponds to a bit-flip occurring with probability $p$ on each qubit. Since \emph{all qubits} experience a local bit-flip channel in our case, this translates into a set of $Q=2^L$ Kraus operators by virtue of Eq.~\eqref{eq:kraus_map}. For single-qubit gates, current technology estimates~\cite{PRXQuantumSondhi2021} of the noise parameter are $p\sim 10^{-3}$.

Further generalizations for the noise models to be employed are possible. We have included the case of a fully depolarizing channel model in Appendix section~\ref{subsec:fdepolm}.

\subsection{Measurement operation}\label{subsubsec:measurement_prot}
A set of $M<L$ qubits forming the domain wall are measured at a given time step $n$ under the action of the $\mathcal{M}$ operation. Calling $x_0^{(n)}$ the coordinate of the left-most qubit forming the domain wall at time step $n$, the domain wall indices at that time step is given by the set of integers:
\begin{eqnarray}\label{eq:preselected_indices}
\mathcal{S}^{(n)}=\{x_0^{(n)},x_0^{(n)}+1,...,x_0^{(n)}+M-1 \}.
\end{eqnarray}
Note that the domain wall $\mathcal{S}^{(n)}$ is dynamic and changes coordinates at any time step $n$ in a random fashion.

We distinguish between two different projective measurement protocols, each represented by projectors $\hat{P}_0$ and $\hat{P}_1$ in the computational basis:
\begin{eqnarray}\label{eq:measurementK}
\hat{P}_0=|0\ra\la 0|,\hspace{10pt} \hat{P}_1 = |1\ra\la 1|.
\end{eqnarray}
To simulate a measurement operation at time step $n$, we proceed as follows: For a selected qubit with index $i\in\mathcal{S}^{(n)}$, we calculate the expectation value of the $z$-component of the Pauli matrix at that particular qubit location $\la\sigma^z_i\ra\in [-1,1]$. We define $m_i=\frac{1-\la\sigma_i^z\ra}{2}$, and generate a random number $r_i\in [0,1]$ from a uniform distribution. If $r_i>m_i$, we apply the measurement protocol with $\hat{P}_0$, simulating a collapse of the qubit over the $|0\ra$ state; otherwise, we employ the protocol with $\hat{P}_1$, simulating a collapse over the $|1\ra$ state. For each disorder realization we then follow the trajectory of measurement outcomes that is determined by the generated random numbers. Note that these operations are non-unitary. The resulting measurement outcomes are recorded into a classical register with $M$ values, where each value being either $+1$ or $-1$ for any index in $\mathcal{S}^{(n)}$. The measurement values are recorded into an $M$-entry vector:
\begin{eqnarray}\label{eq:vec_meas}
\vec{v}\left(n\right) &=&(\sigma_{x_0},\sigma_{x_0+1},...,\sigma_{x_0+M-1})|\sigma_j\in\{+1,-1\},\nonumber\\
\sigma_j &\in&\{+1,-1\}\hspace{5pt} j\in \mathcal{S}^{(n)}.
\end{eqnarray}
Note that the vector depends implicitly on the coordinates of the domain wall. For that particular domain wall indices set, there is an associated vector at initialization, $\vec{v}(n=0)$, which is known from the initial preparation of the product state. The above measurement procedure is repeated at any time step $n$ and with the associated domain wall $\mathcal{S}^{(n)}$. 

\subsection{Correction operation}
\label{subsubsec:correction_protocol}
Here we present the correction operation $\mathcal{C}$. Due to the probabilistic nature of a measurement, the value obtained for a given qubit might be erroneous (i.e. it breaks the subharmonic response pattern characterizing the DTC). To overcome this, the correction protocol $\mathcal{C}$ is applied immediately after the measurement operation $\mathcal{M}$, with $\mathcal{C}$ being solely based on the initial state configuration $\rho_0$ and the obtained measurement outcomes at the time step $n$; we refer again to Fig.(1) (a) for a schematic representation of this operation. 

The initial correlation matrix between qubits at positions given by $\mathcal{S}^{(n)}$ is known at time $n=0$, and it is defined as the outer product of the associated vector at initialization:
\begin{eqnarray}
C_0(i,j)=\vec{v}(0)\otimes \vec{v}(0), \hspace{5pt}i,j\in \mathcal{S}^{(n)}.
\end{eqnarray}
The associated correlation matrix of the recorded outcomes for the domain wall $\mathcal{S}^{(n)}$ at time $n$ is:
\begin{eqnarray}
C_n(i,j)=\vec{v}(n)\otimes \vec{v}(n), \hspace{5pt}i,j\in \mathcal{S}^{(n)}.
\end{eqnarray}
Both matrices, which are the result of a classical computation, are compared by their element-wise product:
\begin{eqnarray}
\delta_{ij}(n)=\text{int}\left(\frac{1}{2}\left(\mathrm{J}_{ij}-C_0(i,j)*C_n(i,j)\right)\right),
\end{eqnarray}
where $\text{int}$ represents the integer part, $\mathrm{J}_{ij}$ is the matrix of ones of dimension $M\times M$, and $*$ indicates element-wise multiplication. Note that $\delta_{ij}(n)$ is a real, symmetric matrix with integer entries. In the case where noise levels are moderate and we are deep in the DTC phase with $g\sim 1$, it is expected that repeated measurements yield entries for $\vec{v}(n)$ that follow the subharmonic pattern characterizing the MBL-DTC phase. Thus, we mask those elements that do not experience a sign change in the components $\delta_{ij}(n)$. This way, the non-zero entries of the matrix correspond to qubits experiencing a sign flip respect to $C_0(i,j)$. We define the position of the qubit to be corrected at that time step $n$ as $i(n)$, given by:
\begin{eqnarray}\label{eq:qimax}
i(n)=\text{max}\left(\sum_j\delta_{ij}(n)\right).
\end{eqnarray} 
Unless otherwise stated, we correct one qubit out of the total $M$ measured qubits forming the domain wall. If more than one qubit is corrected, the indices follow the criteria employed in Eq.~\eqref{eq:qimax}, from bigger to smaller values of $i(n)$.

\subsection{Observables}\label{subsec:observables}
We identify the relevant observables that characterize the MBL-DTC phase following Refs.~\cite{khemani2019brief,PRXQuantumSondhi2021}. We represent a spatially and disordered averaged operator by a double angle bracket. For any operator that can be decomposed as a sum of local terms $\mathcal{O}=\sum_{j=1}^{L}\mathcal{O}_j$, the averaging over $L$ lattice sites and $N_{\text{dis}}$ disorder realizations at stroboscopic times $n$ is given by:
\begin{eqnarray}\label{eq:observables_eq}
\la\la\mathcal{O}(n)\ra\ra &=&\frac{1}{L N_{\text{dis}}}\sum_{j=1}^{L}\sum_{m=1}^{N_{\text{dis}}}\text{tr}\left(\mathcal{O}_j\rho^{(m)}_n\right)\nonumber\\
&\coloneqq &\frac{1}{L N_{\text{dis}}}\sum_{j=1}^{L}\sum_{m=1}^{N_{\text{dis}}}\langle\mathcal{O}_j\rangle_{\rho^{(m)}_n},
\end{eqnarray}
where $\text{tr}$ denotes the trace and $\rho^{(m)}_n$ represents the state of the system for the $m$-th disorder realization of the circuit at time step $n$.

A quantity of experimental relevance~\cite{Zhang2017,Choi2017,DTCexperiment1,Choi2019,Frey2022,Kongkhambut2022,Mi2022} to describe the DTC phase is the spatially and disorder averaged autocorrelation, defined by:
\begin{eqnarray}\label{eq:Cn}
\langle\langle C(n)\rangle\rangle=\langle\langle \sigma^{z}(0)\sigma^z(n)\rangle\rangle.
\end{eqnarray} 

The disorder averaged, equal-time qubit-qubit correlations between qubits at positions $L/2$ and $L$ of the chain is given by:
\begin{eqnarray}\label{eq:szsz}
|\langle \sigma^{z}_{L/2}\sigma^z_L\rangle|(n) = \frac{1}{N_{\text{dis}}}\sum_{m=1}^{N_{\text{dis}}}|\langle \sigma^z_{L/2}\sigma_L^z\rangle_{\rho_n^{(m)}}|.
\end{eqnarray}

A quantity of relevance to determine the existence of glassy spatial patterns is the Edwards-Anderson spin-glass order parameter~\cite{PRXQuantumSondhi2021,Mi2022}:
\begin{eqnarray}\label{eq:edwards_parameter}
    \chi^{\text{SG}}=\frac{1}{L}\sum_{i,j}\langle \sigma^z_i \sigma^z_j\rangle^2.
\end{eqnarray}
This parameter is averaged over the total number of disorder realizations $N_{\text{dis}}$.

The mid-chain entanglement entropy measures the amount of entanglement for a bipartition of the system into two equal size subregions $A,B$. Averaged over disorder realizations, it is given at stroboscopic times $n$ by:
\begin{eqnarray}\label{eq:sent_midchain}
S_{L/2}(n) &=& \frac{1}{N_{\text{dis}}}\sum_{m=1}^{N_{\text{dis}}}\text{tr}(-\rho_{n,A}^{(m)}\ln \rho_{n,A}^{(m)}),\nonumber\\
\rho_A &=&\text{tr}_B \rho.
\end{eqnarray}

For pure states, this quantity can be computed easily by representing the state as a Matrix Product State (MPS) in canonical form~\cite{Vidal2003,Schollwoeck2011} (inner sums indicating contractions are implied):
\begin{eqnarray}
    |\psi\rangle=\Gamma^{\sigma_1}\Lambda_1 ...\Gamma^{\sigma_{L/2}}\Lambda_{L/2} \Gamma^{\sigma_{L/2+1}}... \Lambda_{L-1} \Gamma^{\sigma_{L}}.
\end{eqnarray}
The entanglement entropy of the bipartition in this case is associated with the singular values encoded in $\Lambda_{L/2}$ as:
\begin{eqnarray}\label{eq:sl2_unitary}
    S_{L/2}^{\text{unitary}}(n)=-\text{tr}\left( \Lambda_{L/2}^{2}\log \Lambda_{L/2}^{2} \right).
\end{eqnarray}
For mixed states represented by a MPDO, calculating the entanglement of the bipartition poses a challenging computational task~\cite{GuthJarkovsky2020,Preisser2023}. Since local updates of the MPDO only occur in the  $A^{\sigma_j,r}_{\chi_j\chi_{j+1}}$ in Eq.~\eqref{eq:mpdo}, we associate, in analogy with the unitary case, all updates of singular values in the middle bond of the chain as a measure of quantum correlations in the state, i.e. as a measure to characterize entanglement dynamics in the MPDO. We therefore extend the definition of entanglement to mixed states, with the mid-chain entanglement entropy in Eq.~\eqref{eq:sl2_unitary} corresponding to the limiting case under unitary evolution for pure states. We call this generalization the operator entanglement entropy. We stress that this definition does not necessarily represent the entanglement of the bipartition of the chain in the MPDO case~\cite{GuthJarkovsky2020,Preisser2023}, but rather serves as a proxy to the development of quantum correlations in the state even in the presence of non-unitary operations.

\section{}\label{app:appb}

\subsection{Operator entanglement growth and probability of correction for different initial states}

Following the discussion on the initial state independence of results in the main text, here we report additional data regarding the mid-chain operator entanglement entropy, the probability of correction and the averaged qubit purities, for the four different product states considered in Sec.~\ref{subsec:init_state_indepe}. 

The average amount of entanglement for a single qubit with index $j$ respect to the rest of a quantum system is represented by the \emph{purity} of the reduced density matrix $\rho_{n,j}$ at stroboscopic times $n$, defined by the partial trace over the system degrees of freedom excluding qubit $j$:
\begin{eqnarray}
\rho_{n,j} = \text{tr}_{\Omega,j\notin \Omega}\left(\rho_n\right),
\end{eqnarray}
where $\Omega$ represents the system excluding the qubit at the lattice index $j$. The spatially and disorder averaged qubit \emph{purities} are defined by:
\begin{eqnarray}
\langle\langle P\rangle\rangle(n)=\frac{1}{L N_{\text{dis}}}\sum_{j=1}^L\sum_{m=1}^{N_{\text{dis}}} \text{tr}\left(\left(\rho_{n,j}^{(m)}\right)^2\right).
\end{eqnarray}

\begin{figure}[!ht]
\includegraphics[scale=0.6]{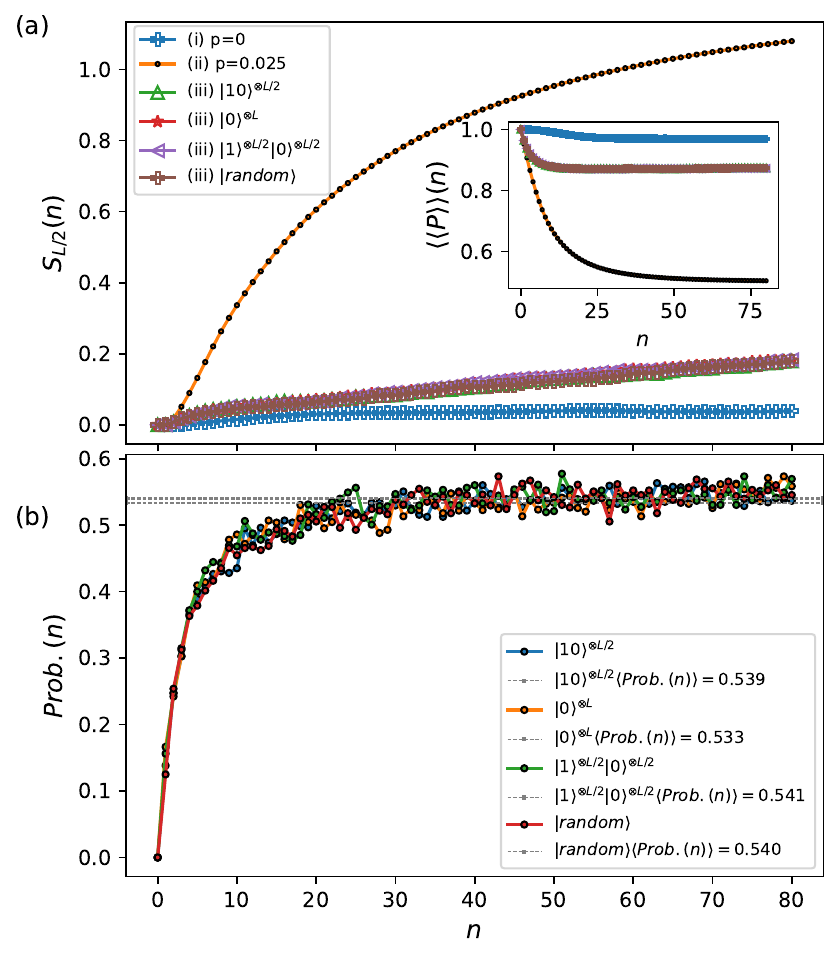}
\caption{{\bf: Operator entanglement and probability of correction dynamics:} (a) The disorder averaged mid-chain operator entanglement entropy $S_{L/2}(n)$. Protocol (i) shows a very slow growth of entanglement compared to the pure dissipative case of protocol (ii). Application of non-unitary operations under protocol (iii) leads to a much slower growth rate for the mid-chain operator entanglement entropy. The inset represents the averaged qubit purities. (b) The probability of applying the correction scheme described in protocol (iii). We observe saturation to a value close to $0.5$ independent of the initial state. All parameters correspond to those of Fig.(2) in the main text.}
\label{fig:figs1}
\end{figure}

In Fig.~\ref{fig:figs1} (a) we represent the mid-chain operator entanglement entropy. We observe nearly identical behaviour of the entanglement growth for the set of selected initial states. Due to the non-unitary operations in protocol (iii), the operator entanglement entropy grows at a much slower rate compared to protocol (ii); we argue that such slow growth could be logarithmic. The (unbounded) logarithmic growth of entanglement is a characteristic feature of quenched product states in one dimensional MBL phases in the thermodynamic limit. In Fig.~\ref{fig:figs1} (b), we have represented the probability of performing a single qubit correction in protocol (iii) at different time steps. The results suggest that this probability is independent of the initially prepared state.

\subsection{Long time behaviour of operator entanglement entropy and probability of correction}

\begin{figure*}[!t]
\includegraphics[scale=0.62]{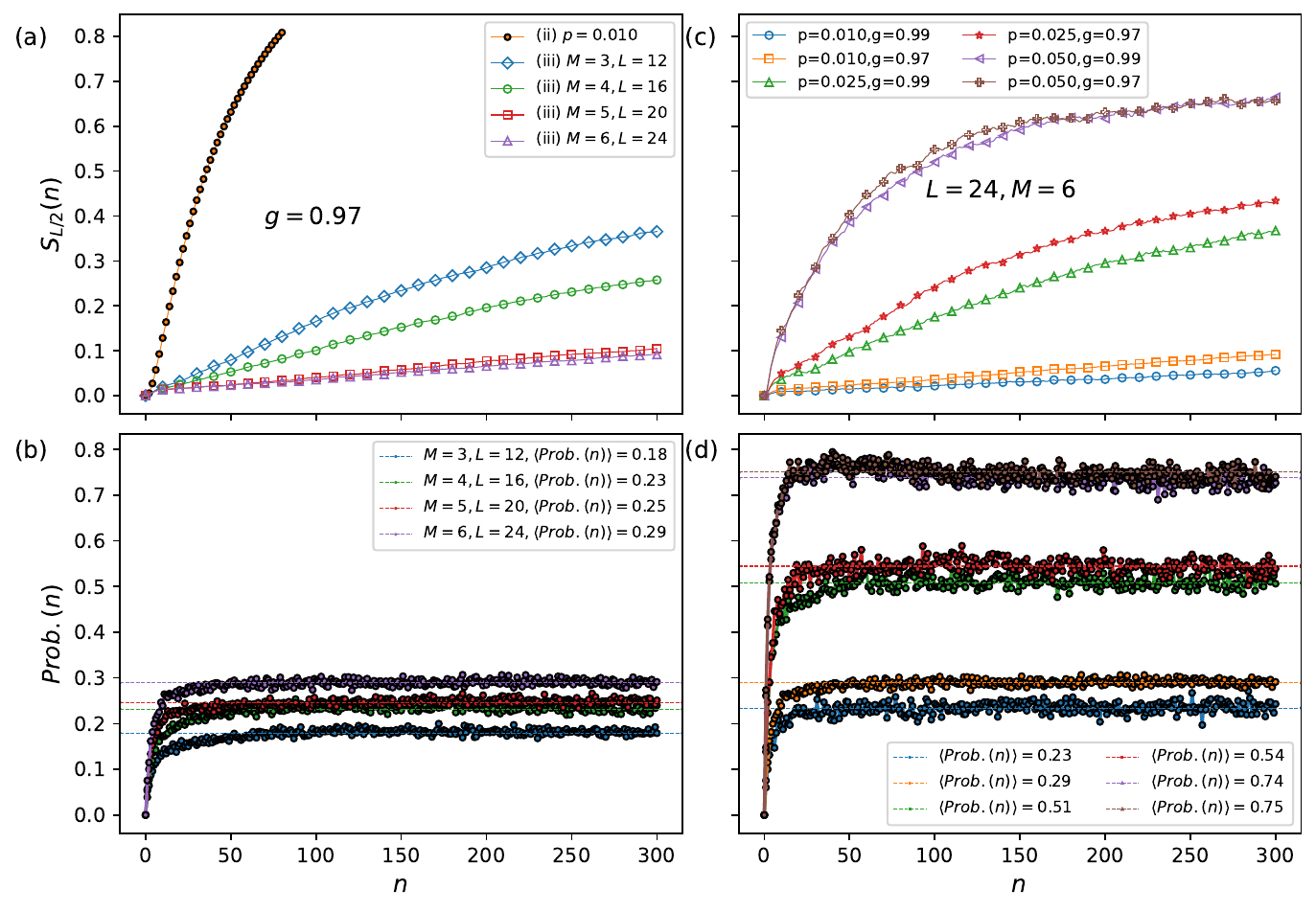}
\caption{{\bf{Long time dynamics of operator entanglement entropy and probability of correction:}} (a) Mid-chain operator entanglement entropy $S_{L/2}$ and (b) the probability of performing a correction under protocols (ii) ($L=24,M=6,N_{\text{dis}}=256$) and (iii) ($N_{\text{dis}}=1280$ and different system sizes), for fixed $p=0.01,g=0.97$. (c) Mid-chain operator entanglement entropy $S_{L/2}$ and (d) the probability of performing a correction under protocol (iii), for different values of $p$ and $g$, fixing $L=24,M=6$.}
\label{fig:figs2}
\end{figure*}

In Fig.~\ref{fig:figs2}. we have represented the mid-chain operator entanglement entropy, along with the probability of performing a correction operation in protocol (iii), for different tuning of the parameters. In Fig.~\ref{fig:figs2}~(a), (b), increasing system size leads to a decrease in the growth rate of the operator entanglement entropy, while at the same time the probability of correction saturates. The behavior of the operator entanglement entropy is in accordance with the behavior observed in the autocorrelator in the main text. In panels (c), (d), the major role played by the noise level $p$ respect to the pulse parameter $g$ is manifest. Colors in (d) apply those in the legend in (c).

\subsection{Fully depolarizing noise model}\label{subsec:fdepolm}
We have in addition tested the protocol under a fully depolarizing noise model, in which each qubit experiences a depolarizing channel given by Kraus operators:
\begin{eqnarray}
K_0 &=& \sqrt{1-p}I,\hspace{5pt} K_{s=1,2,3}=\sqrt{p/3}\sigma^s, \nonumber\\
\sigma^{s=1,2,3}&=&\sigma^{x,y,z}.
\end{eqnarray}
We note that under this definition of the quantum channel, the value of the noise parameter $p$ is not equivalent to the one employed for the bit-flip channel.

\begin{figure*}[!t]
\includegraphics[scale=.39]{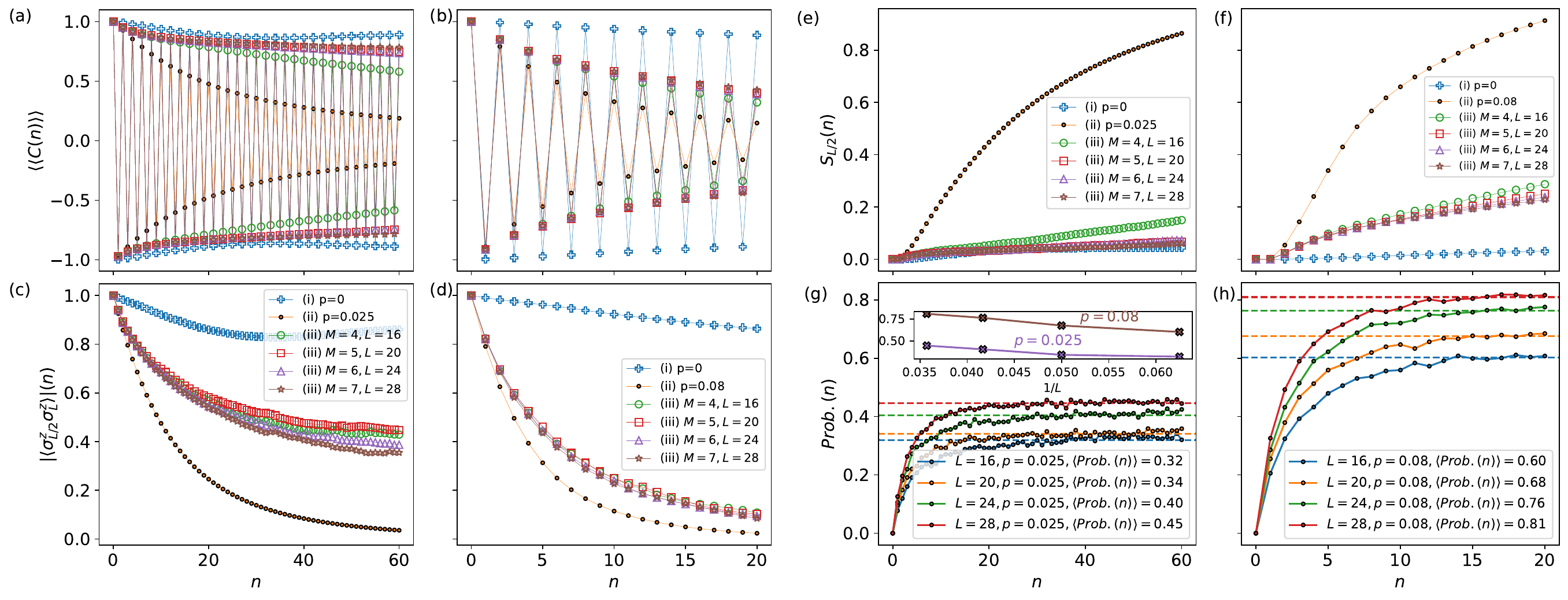}
\caption{Results for the depolarizing noise model in protocols (i), (ii) and (iii), for different system sizes $L$, at noise levels of $p=0.025,0.08$ for $N_{\text{dis}}=5120$ disorder realizations. }
\label{fig:figsdepol}
\end{figure*}

The results are represented in Fig.~\ref{fig:figsdepol}. We observe that the main results remain robust under depolarizing noise channel for the values of noise parameter $p$ reported, with spatial and temporal correlations showing a clear amplification for different system sizes under protocol (iii).  

\subsection{Numerical convergence}\label{subsec:numconv}
Here we provide evidence that variation of the tolerance threshold parameter $\varepsilon$ and the Kraus auxiliary dimension $K$ in the time evolution of the MPDO does not lead to different results (see Appendix~\ref{subsec:mpdo} for a definition of these parameters). 

To test this, we focus on a single circuit realization of the system for $L=24$ qubits, $n_s=60$ and $g=0.97$. The same realization of the circuit is employed in all simulations for variations of both $\varepsilon$ and $K$. For protocol (iii) we must, in addition of using the same circuit realization, employ the same measurements and corrections in all the cases; i.e. for any pair $(\varepsilon,K)$, both the locations of the domain wall at any step $n$ \emph{and} the simulated measurement outcomes are the same. We exclude protocol (i) because this protocol does not depend on the Kraus dimension $K$ due to unitary evolution of the state.

\begin{figure}[!t]
\includegraphics[scale=0.46]{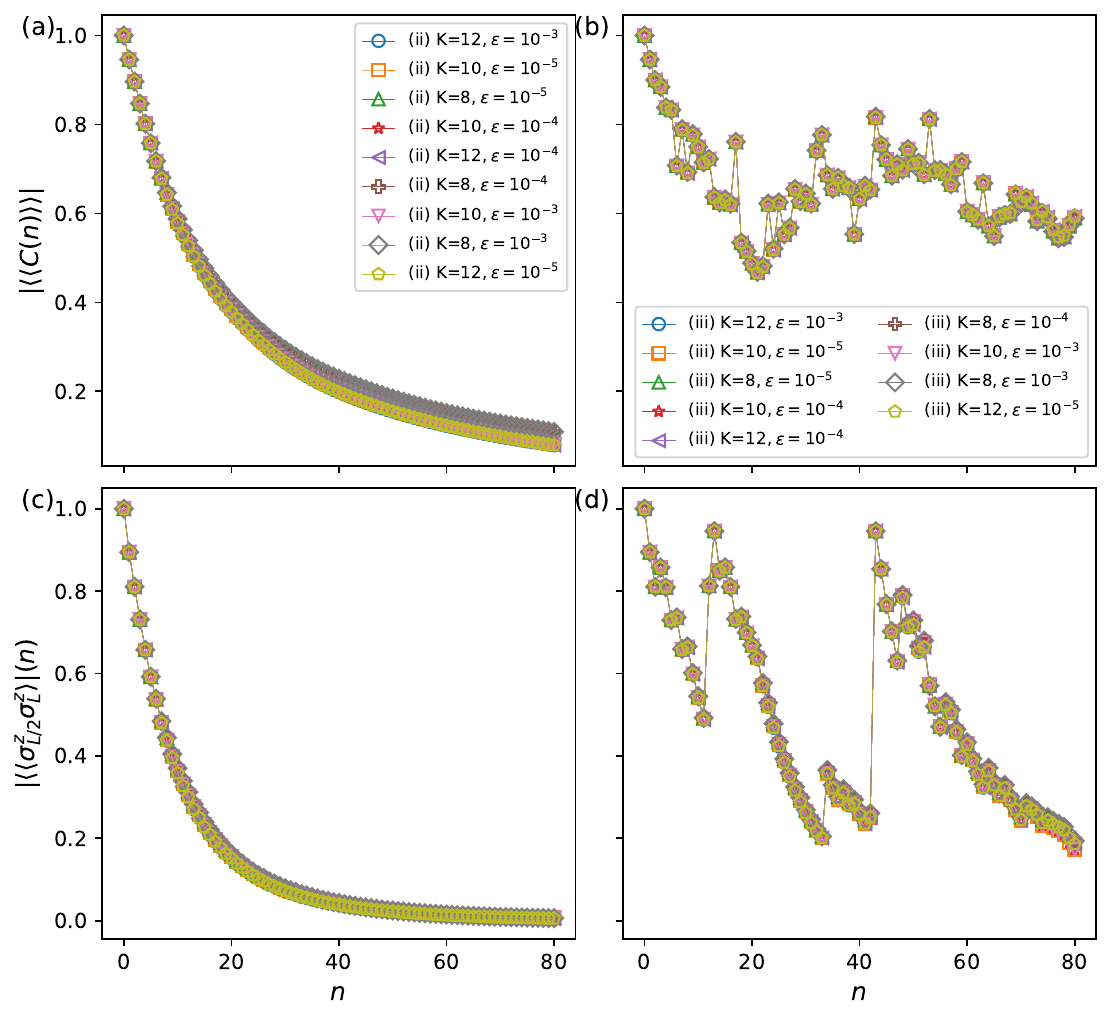}
\caption{Convergence of results for protocols (ii) and (iii), for $L=24,g=0.097,p=0.025,N_{\text{dis}}=1$ for the autocorrelator (panels (a), (b)) and the qubit-qubit correlation (panels (c), (d)), for variations of $\varepsilon$ and $K$ (see Appendix section~\ref{subsec:mpdo}), for a single disorder realization and the same set of measurement outcomes in all cases, as explained in the text. Results reported in the main text correspond to $\varepsilon=10^{-4}$ and $K=10$. The jumps observed under protocol (iii) correspond to the successive application of projective measurements.}
\label{fig:figs3}
\end{figure}

In Fig.~\ref{fig:figs3}, results for the autocorrelator and the qubit-qubit correlation show nearly identical values in all cases, with the exception of $K=8$, which is slightly deviated from the reference case $\varepsilon=10^{-5},K=12$. Note that results reported in the main text correspond to $\varepsilon=10^{-4}$ and $K=10$.

\subsection{Exact evolution benchmark}\label{subsec:edbench}
In order to benchmark the MPDO approach and the application of non-unitary gates as measurements, here we show comparison of the MPDO evolution of the state against the exact evolution of the density matrix operator (ED) for small chain sizes. We fix $L=8$ as our system size, and create a single disorder realization for the couplings. We also create all measurement events at any time step before executing both algorithms, so that the evolution of the state is exactly equivalent under both methods.

The method is benchmarked employing the noise model advertised in the main text (see Appendix section~\ref{subsec:Noise_models}). This noise model applies a mapping over the density matrix $\rho$ with a total of $2^L$ possible outcomes for the state. In the exact simulation, constructing the associated Kraus operators acting on $\mathbb{C}^{2^{L}}$ can be done employing the computational basis states. For a given state in the computational basis, any site is associated with a Kraus operator $K^0=\sqrt{1-p}I$ ($K^1=\sqrt{p}\sigma^x$) for the zero state (one state), where $I,\sigma^x$ are the identity and the Pauli sigma $x$ matrices acting on the site Hilbert space of local dimension $\mathbb{C}^{2}$. Thus, a single outcome $\rho^{\{s\}}$ out of the $2^L$ possibilities is given by:
\begin{eqnarray}
    \rho^{\{s\}}&=&\mathcal{K} \rho \mathcal{K}^\dagger,\nonumber\\
    \mathcal{K}&=&K_1^s\otimes K_2^s\otimes...\otimes K_j^s\otimes...\otimes K^s_L,\nonumber\\
    K_j^s&=&\sqrt{1-p}I_j\lor \sqrt{p}\sigma^x_j.
\end{eqnarray}

\begin{figure}[!t]
\includegraphics[scale=0.46]{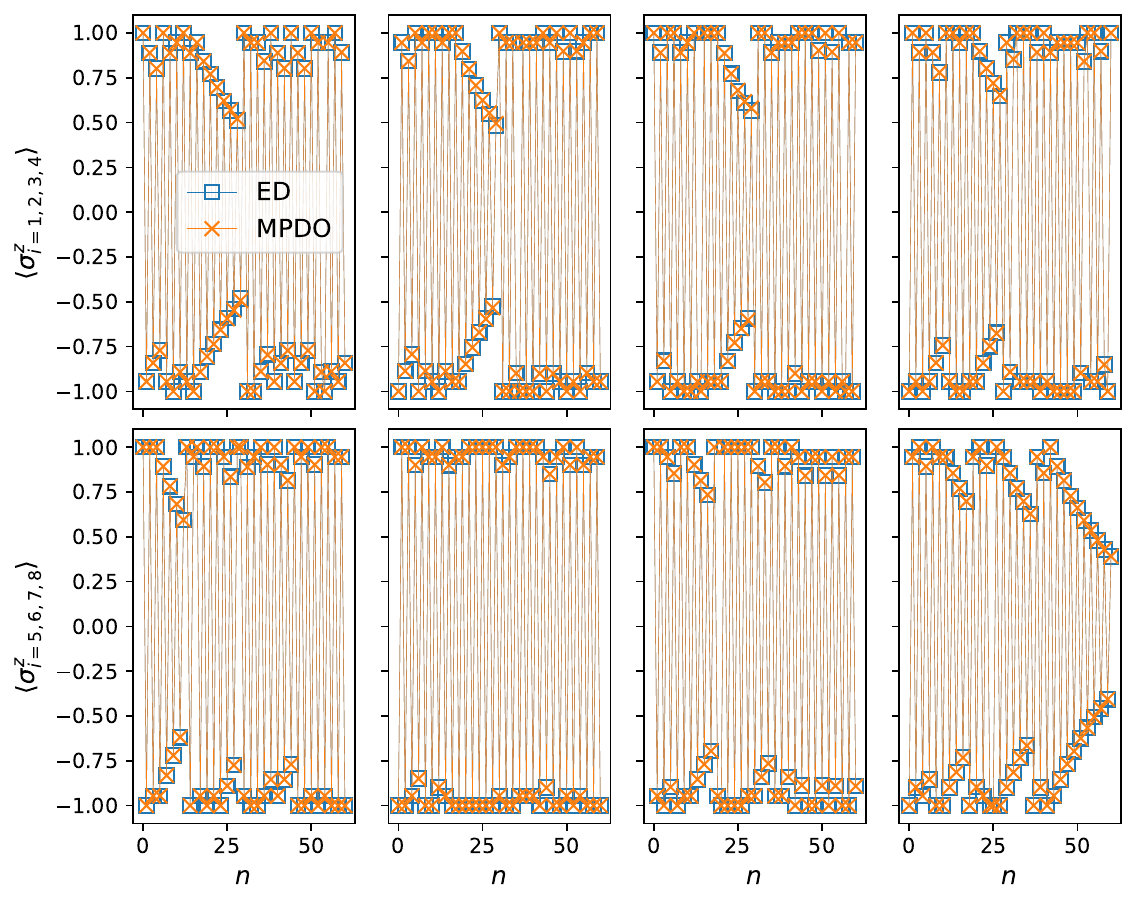}
\caption{Comparison of the MPDO method against the exact evolution of the state for a chain of size $L=8$, with $g=0.97$, $p=0.025$ and $M=3$ for the size of the domain wall. The figure represents the local magnetization $\langle\sigma_i^z\rangle$ at different sites $i$ of the chain.}
\label{fig:figs4}
\end{figure}

\begin{figure}[!t]
\includegraphics[scale=0.46]{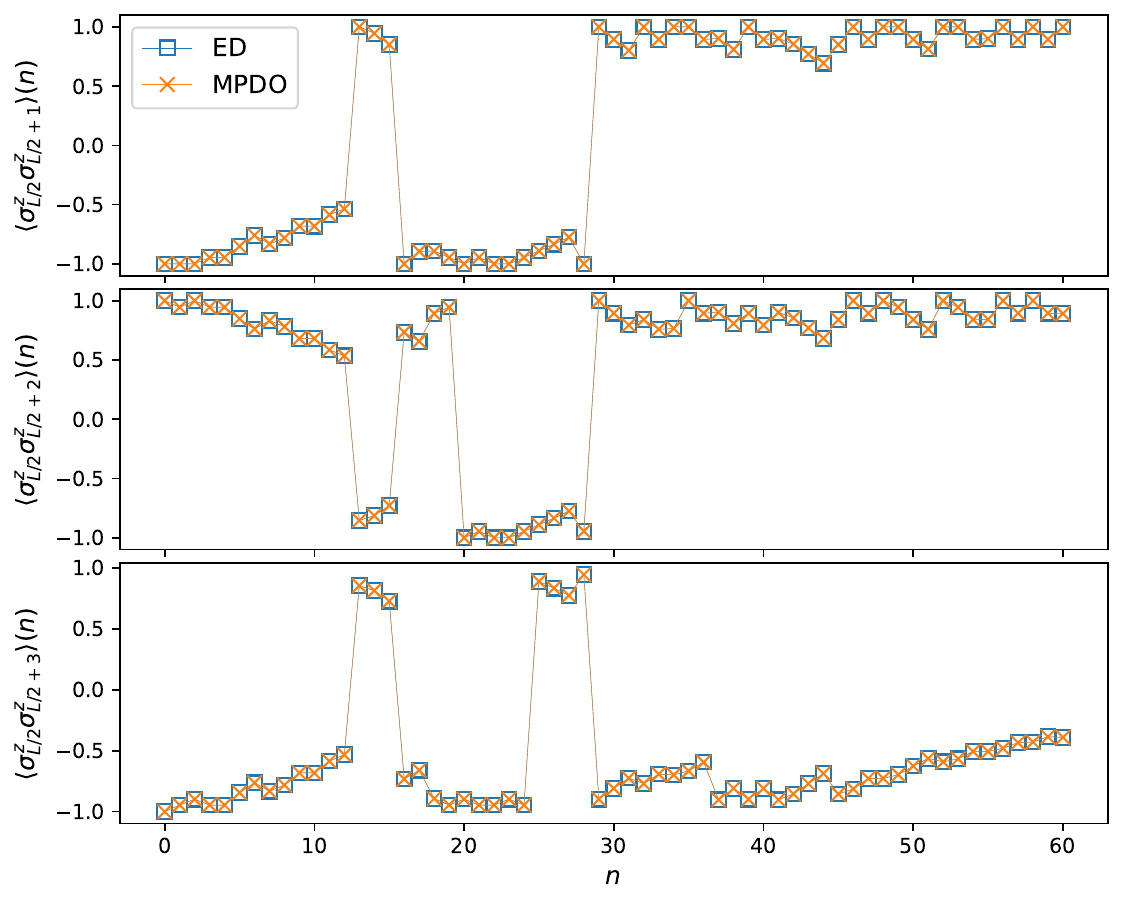}
\caption{Comparison of the MPDO method against the exact evolution of the state for a chain of size $L=8$, with $g=0.97$, $p=0.025$ and $M=3$ for the size of the domain wall. The figure represents the qubit-qubit correlations $\langle\sigma_{L/2}^z\sigma_{L/2+i}^z\rangle$ at different distances $i$ from the middle of the chain.}
\label{fig:figs5}
\end{figure}

The comparison between both methods is shown in Figs.~\ref{fig:figs4}, \ref{fig:figs5}.

\clearpage

\bibliography{refs}
\newpage

\end{document}